# Observations of a new stabilising effect for polar water ice on Mars

Adrian J. Brown[1*1], Jonathan Bapst[2], Shane Byrne[2]

[1] *Plancius Research, Severna Park, MD 21146, USA*
[2]*Lunar and Planetary Laboratory, University of Arizona, Tucson, AZ, 85721, USA*

**Using the Compact Reconnaissance Imaging Spectrometer for Mars (CRISM), we map the temporal variability of water ice absorption bands over the near-polar ice mound in Louth crater, Mars. The absorption band depth of water ice at 1.5 microns can be used as a proxy for ice grain size and so sudden reductions can time any switches from ablation to condensation. A short period of deposition on the outer edge of the ice mound during late spring coincides with the disappearance of seasonal water frost from the surrounding regolith suggesting that this deposition is locally sourced. The outer unit at Louth ice mound differs from its central regions by being rough, finely layered, and lacking wind-blown sastrugi. This suggests we are observing a new stabilizing effect wherein the outer unit is being seasonally replenished with water ice from the surrounding regolith during spring. We observe the transport distance for water migration at Louth crater to be ~4km, and we use this new finding to address why no water ice mounds are observed in craters <9km in diameter.**

Key Point 1. First observation of Martian process of exchange of perennial water with the surrounding regolith
Key Point 2. We infer a advection travel distance of ~4km for water ice particles
Key Point 3. Our findings could explain why no craters <9km have water ice mounds

Corresponding author:
Adrian Brown
Plancius Research
ph. 408 832 6290
email: Adrian.J.Brown@nasa.gov

---
[1]

## 1. Introduction

The North Polar Residual Cap (NPRC) is a surface water ice deposit that persists throughout the martian summer (Kieffer 1976) and which unconformably drapes the much-thicker North Polar Layered Deposits (Byrne 2009; Herkenhoff et al. 2007; Tanaka et al. 2008). The NPRC is extensive (~$10^6$ km$^2$) and plays a central role as the dominant source and sink of seasonal water in the martian atmosphere (Smith et al. 2016). Analysis of its albedo (Kieffer 1990) and spectra (Langevin et al. 2005; Brown et al. 2016) show the NPRC composition is dust-free large-grained water ice.

Multiple impact craters peripheral to the NPRC contain surface ice with similar appearance, albedo and spectra (Conway et al. 2012; Brown et al. 2008). Korolev crater contains the largest of these deposits and has thermal properties (Armstrong et al. 2005) and underlying layering (Brothers and Holt 2016) that is similar to the NPRC. Louth crater contains the most equatorial of these deposits and has been identified as similar in surface texture and composition to the NPRC (Brown et al. 2008).

Water ice is seasonally added and removed from these deposits in a repeatable way. Each winter, atmospheric water in the north polar hood is incorporated into the seasonal frost cap. As this seasonal cap retreats water is liberated from its edge; however, most of this water recondenses on the remaining seasonal ice immediately to the north as originally described by Houben et al. (1997). This Houben-effect results in an annulus of fine-grained water frost tracking the edge of the seasonal cap as it retreats poleward (Wagstaff et al. 2008), which sweeps over craters like Louth and Korolev in the late Spring (Byrne et al. 2008; Brown et al. 2012). During summer the seasonal frosts disappear and large-grained (old) water ice appears as the NPRC is exposed. Brown et al. (2016) used CRISM to measure 1.5 μm adsorption band depth and so identify when fine-grained frost began accumulating in late summer at Gemini Lingula ($L_s$~110-130). This reaccumulation causes late summer brightening of the NPRC observed by Mariner and Viking IRTM (James and Martin, 1985; Kieffer, 1990; Paige et al., 1994; Bass et al., 2000), TES (Armstrong et al., 2007; Xie et al., 2008) and MEX-OMEGA (Appere et al., 2011).

The net annual mass balance of the NPRC and its outliers is uncertain. The exposure of larger-grained (old) ice in the summer indicates that net ablation is taking place; however, the lack of surface dust indicates that ablation has been minimal so it's possible that the NPRC has only recently entered this state (Laskar et al. 2002; Greve et al. 2005). At 70.2N, Louth crater contains the lowest latitude surface ice on Mars and so it may be even more sensitive to any recent climatic changes that may have occurred.

Here we study the Louth crater ice mound using the same techniques as Brown et al. (2016) applied to the NPRC. This convex-up dome-shaped mound is 10 km wide, 250 m thick and displays layering (Conway et al., 2012). Several formation mechanisms have been proposed for this mound and others like it, including 1) remnants of a more extensive northern ice cap (Garvin et al., 2000; Tanaka et al., 2008), 2) upwelling from an underground aquifer (Russell and Head, 2002), 3) melting and ponding of near surface ice by impact induced hydrothermal activity (Rathbun and Squyres, 2002) and 4) atmospheric direct deposition as individual outliers (Brown et al., 2008; Conway et al., 2012). We examine the seasonal timing of switches between accumulation and ablation, compare these observations to recent simulations of Bapst et al. (2017) and discuss the implications of our results for the formation of the ice mound and its long term fate.

## 2. Methods

CRISM is a visible to near-infrared imaging spectrometer with 545 channels from 0.365-3.94 μm (Murchie et al., 2007). In this work, we utilize Full Resolution Targeted (FRT: 545 channels, 18 m/pix), Half resolution long (HRL: 545 channels, 36 m/pix) and Multispectral Mapping (MSP: 72 channels, 180m/pix) observations. CRISM observations are ~10km wide, but vary in length by observation type.

The depth of the 1.5 μm water ice absorption band is a non-linear proxy for ice grain size (Langevin et al. 2005; Brown et al., 2012, 2014, 2016) and is defined by an $H_2O$ index:

$$H_2O\,index = 1 - \frac{R(1.5)}{R(1.394)^{0.7} R(1.75)^{0.3}} \qquad (1)$$

where $R(x)$ is the reflectance at wavelength $x$. This index saturates at values of ~0.7 or grain sizes of ~100 μm. Increases in the $H_2O$ index (grain size) may occur by removal or thermal metamorphism of fine-grained ice, whereas decreases are commonly associated with condensation or precipitation of fresh finer-grained water ice.

We calculated the $H_2O$ index for each CRISM pixel and used the "MR PRISM" software suite (Brown et al., 2004; Brown et al., 2005; Brown and Storrie-Lombardi, 2006) to map-project these images into a polar stereographic space with a common pixel size.

## 3. Results

Figure 1 shows three CTX images of Louth Crater from late northern summer in Mars Year (MY) 30. The central and right images show late summer images with low albedo markings on the ice mound (red arrows). Similar dark markings that have a streaky appearance have also appeared on the pole facing crater rim (blue arrows).

*3.1 $H_2O$ index maps*

The $H_2O$ index for four periods during mid northern summer in MY29, 30 and 31 are shown in Figure 2a-c, to highlight changes in the water ice index over the Louth ice mound during this period. Figures 2a-c show water ice index images with a stretch that is designed to show small changes in the index on the lower end of the scale. This has the effect of bringing out subtle changes in water ice index in the regolith around the Louth ice mound. This stretch also allows us to observe that during the period, following the retreat of the seasonal $CO_2$ ice cap, there is a large halo of water ice on the crater floor, surrounding the ice mound, which is either deposited during the previous fall before the $CO_2$ seasonal cap was deposited, or was left over as a residuum from the seasonal cap retreat due to the higher stability of water ice compared to $CO_2$ ice (Titus, 2005; Titus et al. 2008; Wagstaff et al., 2008; Brown et al., 2012). A key question we can now address is what is the fate of this regolith-covering water ice as temperatures warm?

Figure 3 displays the MY30 dataset with a stretch that is intended to bring out changes on the high side of the index (corresponding to large grain sizes). This shows the increase in $H_2O$ index, particularly in the $L_s$=108 image, in the center of the mound. Changes around the edge of the mound are much more muted (they appear to be consistently green), however the white arrows indicate the $H_2O$ index on the southern edge of the mound increases significantly from $L_s$=59 to 108.

The five $H_2O$ index maps for MY29 covering the $L_s$=62-149 period are shown in Figure 2a. These show the gradual increase of the $H_2O$ index over the ice mound over this time. The last image at $L_s$=149 is almost uniformly white (indicated by a pink arrow). The images also show a decrease in the $H_2O$ index in the regolith around the ice mound. In particular, it should be noted that at $L_s$=62-67, around the southern rim of the ice mound, a halo of water ice is present (this is colored light blue and indicated by a white arrow in the $L_s$=62 image). This disappears markedly in the $L_s$=92-149 period. The last two images show strong $H_2O$ index values in an annulus which is the rim of the crater to the south of the ice mound.

The MY30 $H_2O$ ice index maps are shown in Figure 2b over the period from $L_s$=59, when seasonal $CO_2$ is finally disappearing, to $L_s$=108. It should be noted that the first image at $L_s$=59 shows a relatively high $H_2O$ index on the regolith around the edge of the cap (this is colored green and is indicated by a white arrow). This large ice index is largely gone by the $L_s$=83 image, similar to the previous Mars Year. This is colored red and indicated by a green arrow on the $L_s$=83 image.

The MY31 $H_2O$ index images covering $L_s$=75-165 are shown in Figure 2c. This image sequence again shows an increase in $H_2O$ index on the interior of the mound. The possible exception is for the late summer $L_s$=92 image, which appears to have a slightly lower $H_2O$ index on the north edge of the water ice mound (indicated by a pink arrow in the image). We do not have a satisfactory explanation for this observation. As for the previous two years, the first image shows the highest $H_2O$ index on the regolith surrounding the ice mound. The $L_s$=75 image shows a spatial variation in the water ice index, being lower around the edge of the mound (this is indicated by a white arrow).

Figure 3 displays the MY30 dataset with a stretch that is intended to bring out changes on the high side of the index (corresponding to large grain sizes). This shows the increase in $H_2O$ index, particularly in the $L_s$=108 image, in the center of the mound. Changes around the edge of the mound are much more muted (they appear to be consistently green), however the white arrows indicate the $H_2O$ index on the southern edge of the mound increases significantly from $L_s$=59 to 108.

*3.3 Difference images*

We constructed difference images for four of the images of Figure 2a, and these are shown in Figure 4a. Only the overlapping pixels in each image are processed and presented. Co-registration of the images was carried out as described in Section 2.1.1. The left panel of Figure 4a shows a slight decrease in the $H_2O$ ice index between $L_s$=65 and 67, colored in orange and indicated by a pink arrow. The right panel of Figure 4a shows a dark rim around the edge of the crater, indicating a large decrease in the $H_2O$ index around the rim. The affected area is colored in black and indicated by a white arrow. The other feature of note is a speckle of blue pixels to the south of the crater in the regolith. These areas also experienced a decrease in the $H_2O$ index between $L_s$=67 and 92.

Figure 4b shows the three difference images between the three images of Figure 2b calculated in the same manner as Figure 4a. Most strikingly, Figure 4b shows that in $L_s$=59-83, there is a decrease in $H_2O$ ice index in the regolith surrounding the ice mound (these areas are colored green and blue and indicated by a pink arrow), particularly on the south side of the mound. In the period following this (from $L_s$=83-91), there is a corresponding increase in the $H_2O$ index on the south side of the ice mound (this region is colored white and the location indicated by grey arrows). The spatial occurrence of the increased $H_2O$ index on the south side of the mound and decrease in the water ice in the adjacent regolith in the following period are significant and the possible causes of this observation will be addressed in the Discussion section below.

Comparing Figure 4a and 4b, we can make a direct comparison between the second panel of 4a ($L_s$=67-92) and the first panel of 4b ($L_s$=59-83). The time periods are slightly different, which makes a significant difference, and in addition to this interannual differences will play a confounding role here. However, there are some similarities which lead us to believe the behavior we observe is a regular occurrence. For example, the interior of the water ice mound is white in both images, indicating increasing of $H_2O$ index in this region. The $H_2O$ ice index of the surrounding regolith is blue in some locations, indicating a decrease in those regions. The rim of the mound in Figure 4a (MY 29 $L_s$=67-92) is black, indicating a strong decrease in the $H_2O$ index along the rim at this time, in addition, the rim of the mound in Figure 4b (MY 30 $L_s$=59-83) changes from white to red, indicating a small decrease at that time. Therefore we believe similar trends exist in this key period in MY 29 (Figure 4a) and MY 30 (Figure 4b).

*3.4 Average and maximal $H_2O$ ice index*

Figure 5 shows a gridded spatial and temporal sequence that captures each the $H_2O$ index maps of each CRISM image that obtained useable data over the Louth ice mound (see also Table 1). The images are arranged as function of time in solar longitude (from $L_s$=59 to 149, left to right) for Mars Year 29, 30 and 31, from top to bottom. Note that the temporal (left-right) scale is not linear, we have simply ordered the images from left to right and attempted to line up images of the same time in a Mars Year.

Using this gridded time sequence, we can more easily perceive three points that are critical to establishing the intra and inter-annual behavior of ice on the Louth ice mound. First, the amount of water ice on the surrounding regolith decreases from $L_s$=59 to $L_s$=165, having reached a stable minimum amount by about $L_s$=92.

The second point to note is that the water ice index on the water ice mound interior increases relatively steadily over the first two Mars Years (29, 30), however in one $L_s$=80 observation in Mars Year 31, the water ice index increases and then decreases in $L_s$=92. This phenomenon is not as strongly apparent in the other Mars Years. We address this point further in the Discussion section below.

The third major point to note is the decease in the water ice index around the edge of the mound, particularly on the north side around $L_s$=92 in each Mars Year (this is colored red and indicated with green arrows) but also on the south side of the ice mound (indicated by pink arrows). These decreases are likely caused by deposition of fine grained water ice around the edge of the ice mound – this observation is critical to our model, and is addressed further in the Discussion section.

The late summer CRISM image MSP D25A at $L_s$=149 shows the highest $H_2O$ index values (this image is also shown in Figure 3a). There is an indication of a turn over in the MSP 314FD observation of MY30 at $L_s$=165 (see also Figure 2c), although this could also be due to the onset of the polar hood obscuring the crater at this time (Benson et al., 2011; Brown et al., 2016).

*3.5 HiRISE spring and summer images of Louth*

In order to further our understanding of the regolith-mound water ice exchange process, Figure 6 shows two HiRISE images taken at $L_s$=75.1 and $L_s$=156. The close up images in the bottom of the image show the rough "stucco" nature of the exterior unit of the mound. This was stucco unit was mapped as "Unit 4" by Brown et al. (2008). Two key morphological observations are important to note in these HiRISE images. The first is that the northern rim of the ice mound contact with the regolith is clean and relatively distinct (blue arrows in Figure 6). The southern rim of the ice mound is ragged and less distinct (red arrows). This is a strong indication of southerly winds transporting water ice in the northerly direction, because the ragged edge indicates the direction from which ice is removed, and the sharp edge shows the direction in which ice has newly covered regolith. This is in accord with the asymmetric offset of the ice mound within Louth.

The second observation to make regarding these HiRISE images is the linear features in the middle of the central ice mound. These were interpreted to be "sastrugi" by Brown et al. (2008) and found to be present in their "Unit 1, the smooth undulating layer in the center of the Mound", and less prominently in their mapped "Unit 3". These features run almost north-south (see black lines in Figure 6), which is in accord with their interpretation as wind-blown sastrugi, which on Earth run parallel to the wind direction (Gray and Male, 1981). This is roughly to the north-north east, in accord with the sharp (blue arrows) and ragged (red arrows) edges discussed above.

**4. Discussion**

*4.1 Seasonal evolution of water frosts*

It is clear from Figures 4 and 5 that the seasonal behavior of the mound is not uniform and is quite different from the surrounding regolith. The water ice index over the regolith decreases markedly from ~0.4 to ~0.1 during $L_s$=59 to 92 (white arrows in Figure 2a and b), and then stabilizes at these low values (dark blue in figures) indicating the disappearance of a fine-grained seasonal water frost. In contrast, the water ice index on the ice mound generally starts higher (~0.65 at $L_s$=80 in MY31) and then undergoes a slight decrease (to ~0.6) at the edges of the ice mound from $L_s$=83-91 (green and pink arrows in Figure 5). Thus, the timing of this decrease in water ice index at the mound edge corresponds to the removal of seasonal water frost on the surrounding regolith.

We suggest that water sublimated from the regolith recondenses on the periphery of the (much colder) mound. The spatial pattern of this late-spring condensation is crescent-shaped (Figure 4a), being wider on the southwest edge of the mound. Winds from the southwest could advect regolith-derived water vapor over the mound. As vapor passes over this colder surface it condenses (lowering the water ice index); however, after traversing ~4km of this icy surface the humidity of the air has fallen to the point where further condensation does not occur (leaving the mound interior frost-free). This is similar to the 'Houben' effect that plays out at larger scales at the edge of the retreating seasonal cap (Houben et al. 1997; Wagstaff et al. 2008). Similar mechanisms have also been suggested to create albedo patterns on the North Polar Residual Cap that could ultimately lead to the formation of troughs (Ng and Zuber, 2006).

In addition to its differing water ice index evolution, the periphery of the mound exhibits distinct geomorphology. Figure 9 of Brown et al. (2008) mapped four different units within the water ice using CRISM and HiRISE observations, including an outer finely-layered 'stucco rough ice' unit and an inner, sastrugi-bearing 'smooth interior unit'. The outer unit corresponds to the area that experiences recondensation of regolith-supplied water vapor and, over time, this additional seasonal frost exchange may lead to its distinct morphology.

*4.2 Interannual comparisons*

Brown et al. (2008) identified small ice outliers in HiRISE imagery next to the edge of the Louth ice mound and concluded that it may be in retreat. A multi-year comparison of the edge location of the ice mound showed no detectable net change however (Bapst et al. 2017); although, some interannual variability was seen.

Figure 5 shows a significant difference in the $H_2O$ ice index between MY 29 and 30. There was a global dust storm in winter of MY 28 (Kass et al., 2007) when Louth crater was covered with seasonal $CO_2$ ice. The dust storm transiently warmed the Martian climate system, significantly affected the late-summer appearance of the south polar residual cap in MY28 (Becerra et al., 2015), and likely led to faster springtime recession of the northern seasonal cap in MY 29 (Piqueux et al., 2015). The unusually low $H_2O$ index during spring of MY29 may be linked to the climatic disruption caused by this dust storm.

*4.3 Comparison to 1D depositional model results*

Bapst et al. (2017) discuss the details of a 1-D depositional model to constrain the mass balance of the ice mounds at Louth and Korolev. This model approximates the Louth Crater ice mound as an infinite ice sheet with a constant water-ice albedo that can be varied for different runs. The results are therefore most applicable for the interior of the mound and do not capture the edge water ice exchange effects discussed (e.g. in Figure 5). Figure 7 shows the results of this model for Louth, for a range of albedo (0.35-0.43), and for the nominal windspeed (3 m s$^{-1}$). For the albedo range shown, the model predicts a short period of accumulation of water ice, followed by a longer period of sublimation during northern summer. This range of models all result in net-annual ablation, with ranging from 3.6 to 0.1 mm of water ice lost annually, assuming an ice density of 920 kg m$^{-3}$.

Of special relevance to this paper is the short period of accumulation of water ice after the $CO_2$ seasonal cap has disappeared, occurring from $L_s$ 60-90, and interpreted as cold-trapping by the water ice mound. Prior to $L_s$ ~80, temperatures are too low to result in net sublimation of water ice. However, during northern summer, the ice mound undergoes net sublimation. The observed transition from accumulation to ablation, on the outer part of the mound, occurs between $L_s$ 92-100 in MY29, $L_s$ 83-91 in MY30 and $L_s$>92 in MY31 (Figure 5). From these constraints, we estimate the timing of this transition between $L_s$ 83-100 and that the mound continues to ablate for the remainder of the summer. The timing of this transition is matched by models with water ice albedos of 0.37-0.41 (Figure 7). The modeled albedo is not necessarily representative of the actual ice deposit, but shows the sensitivity of this parameter and how it can be tested against CRISM observations of water-ice index.

Using this albedo range of 0.37-0.41, the net-annual change predicted from the model of Bapst et al. (2017) are 2.4 to 0.7 mm of water ice loss per Mars year. Thus, the model (constrained by early summer observations) and the late-summer CRISM and CTX data paint a consistent picture of the ongoing net-annual loss of ice from the Louth crater mound.

*4.4 Implications for NPRC and origin of outlying water ice mounds*

The observations of water ice transport to the Louth ice mound from the surrounding regolith in the $L_s$=59-91 period (Figure 2b) may have implications for the NPRC and water ice outliers around the NPRC, in addition to other crater hosted ice mounds such as those in Korolev and Dokka craters. Figure 8 shows an interpretive diagram of the transfer process through spring suggested by our observations. This diagram shows the following key events in the model: 1) $L_s$=0-55, Louth Crater ice mound is covered by seasonal $CO_2$ ice and is inactive. 2) $L_s$=55-70, Water ice is exposed on the regolith surrounding the ice mound. 3) $L_s$=80-90, Water ice transfers from the regolith to the mound in the Louth crater situation. A similar process would also likely take place at the edge of large ice mounds craters like Korolev, other water ice outliers and also the edge of the NPRC.

We find it intriguing that there is an apparent advection distance for water transport from the regolith to the Louth ice mound of ~4km. If this transport distance observed at Louth is applicable to other crater hosted ice mounds, we can use this information to speculate on a previously unexplained polar phenomenon – that no craters less than 9km in diameter host water ice deposits (Conway et al., 2012). Since the travel distance of water by advection in the Louth crater environment is approximately 4km, which is remarkably similar to the observed lower limit of 9km for craters with ice mounds. This may indicate that a protected zone like a crater may have to have a radius that exceeds at least twice this distance (8km) in order for this process to be form and preserve this type of perennial ice deposit.

Our observation of an apparent advection distance of ~4km could suggest that for ice mounds to form (and perhaps remain stable), the ice mounds require a certain size and amount of "free space" in which they can interact with the surrounding regolith and cold trap advected water them that can be transported over km size distances. In the case of Louth, the ice mound is probably protected by the exterior smooth unit that, as we have reported, is fed and likely dominated by regolith bound water. This might be the key to its relatively large size and stability, in spite of its distance from the pole.

Considering the scenario of a smaller crater, one might envision that the water ice that is deposited in the bottom of it during winter might travel kilometers by this spring advection transport mechanism to the crater rims and get trapped there, rather than forming a central mound. This may help explain why no ice mounds have formed in craters <9km in diameter.

This potential explanation is speculative at this stage and requires numerical studies and further polar observations. One powerful observational method would be mapping by an active multispectral lidar, with polarization capability, to discriminate cloud composition, aerosols and dust, as recently discussed in (Brown et al., 2015) and highlighted in the recent study by Heavens (2017).

### 5. Conclusions

We have used CRISM data to map the spatial and temporal behavior of water ice within Louth crater. Data from three Mars years show a seasonal increase of the $H_2O$ index in the interior of the mound and a decrease on the surrounding regolith.

The significant findings of our study are:
1. We have established for the first time that the Louth crater ice mound ablates throughout summer, at least until $L_s$=150 (Figure 5).
2. We established that the outer edge of the Louth water ice mound exchanges water ice with the surrounding regolith. As the region warms in spring, water ice sublimates from the regolith surface and is then cold trapped as fine-grained ice onto the edge of the water ice mound. This fine-grained water ice exchange may contribute to the fine layering observed by Conway et al (2012) and the texture of the external unit of the ice mound identified by Brown et al. (2008).
3. The spatial pattern of frost formation on the mound suggests that southwesterly winds are responsible for carrying water vapor across the mound leading to a ~4km wide zone of temporary frost formation. This finding may play a role in explaining the observation that no craters <9km in diameter posses water ice mounds.
4. The surrounding regolith is usually neglected in 1D models when calculating

stability of exposed surface ice on Mars (Bapst et al. 2017; Dundas and Byrne 2010). This work shows that surrounding regolith can act as an enhanced source of water vapor in some seasons and should be included in future assessments of ice stability.

Evolution of the North Polar Cap of Mars as Observed by OMEGA/Mars Express. Science 307, 1581–1584.

Laskar, J., Levrard, B., Mustard, J.F., 2002. Orbital forcing of the martian polar layered deposits. Nature 419, 375–377.

Murchie, S., Arvidson, R., Bedini, P., Beisser, K., Bibring, J.-P., Bishop, J., Boldt, J., Cavender, P., Choo, T., Clancy, R.T., Darlington, E.H., Des Marais, D., Espiritu, R., Fort, D., Green, R., Guinness, E., Hayes, J., Hash, C., Heffernan, K., Hemmler, J., Heyler, G., Humm, D., Hutcheson, J., Izenberg, N., Lee, R., Lees, J., Lohr, D., Malaret, E., T., M., McGovern, J.A., McGuire, P., Morris, R., Mustard, J., Pelkey, S., Rhodes, E., Robinson, M., Roush, T., Schaefer, E., Seagrave, G., Seelos, F., Silverglate, P., Slavney, S., Smith, M., Shyong, W.-J., Strohbehn, K., Taylor, H., Thompson, P., Tossman, B., Wirzburger, M., Wolff, M., 2007. Compact Reconnaissance Imaging Spectrometer for Mars (CRISM) on Mars Reconnaissance Orbiter (MRO). J. Geophys. Res. 112, E05S03, doi:10.1029/2006JE002682.

Ng, Felix S. L., and Maria T. Zuber. "Patterning Instability on the Mars Polar Ice Caps." *Journal of Geophysical Research: Planets (1991–2012)* 111, no. E2 (February 1, 2006). doi:10.1029/2005JE002533.

Paige, D.A., Bachman, J.E., Keegan, K.D., 1994. Thermal and albedo mapping of the polar regions of Mars using Viking thermal mapper observations: 1. North polar region. J. Geophys. Res. 99, 25959–25991.

Piqueux, S., Kleinböhl, A., Hayne, P.O., Kass, D.M., Schofield, J.T., McCleese, D.J., 2015. Variability of the martian seasonal $CO_2$ cap extent over eight Mars Years. Icarus 251, 164–180.

Putzig, N.E., Phillips, R.J., Campbell, B.A., Holt, J.W., Plaut, J.J., Carter, L.M., Egan, A.F., Bernardini, F., Safaeinili, A., Seu, R., 2009. Subsurface structure of Planum Boreum from Mars Reconnaissance Orbiter Shallow Radar soundings. Icarus 204, 443–457.

Rathbun, J.A., Squyres, S.W., 2002. Hydrothermal systems associated with martian impact craters. Icarus 157, 362–372.

Richardson, M.I., Wilson, R.J., 2002. A topographically forced asymmetry in the martian circulation and climate. Nature 416, 298–301.

Russell, P.S., Head, J.W., 2002. The martian hydrosphere/cryosphere system: Implications of the absence of hydrologic activity at Lyot crater. Geophys. Res. Lett. 29, art. no.-1827.

Smith, I.B., Putzig, N.E., Holt, J.W., Phillips, R.J., 2016. An ice age recorded in the polar deposits of Mars. Science 352, 1075–1078. doi:10.1126/science.aad6968

Tanaka, K.L., Rodriguez, J.A.P., Skinner Jr., J.A., Bourke, M.C., Fortezzo, C.M., Herkenhoff, K.E., Kolb, E.J., Okubo, C.H., 2008. North polar region of Mars: Advances in stratigraphy, structure, and erosional modification. Icarus, Mars Polar Science IV 196, 318–358. doi:10.1016/j.icarus.2008.01.021

Titus, T.N., 2005. Thermal infrared and visual observations of a water ice lag in the Mars southern summer. Geophys. Res. Lett. 32, doi:10.1029/2005GL024211.

Titus, T.N., Calvin, W.M., Kieffer, H.H., Langevin, Y., Prettyman, T.H., 2008. Martian polar processes, in: Bell, J.F. (Ed.), The Martian Surface: Composition, Mineralogy, and Physical Properties. Cambridge University Press, pp. 578–598.
~ 13 ~

## 7. Acknowledgements

This work was sponsored by NASA MDAP Grant number NNX16AJ48G. We would like to acknowledge the work of the CRISM science operations team at JHU APL who acquired this wonderful dataset. We would like to thank two anonymous reviewers for their helpful comments. We would also like to thank the creators of *The Expanse* for key inspirations to this project, and thank in advance the future members of the MCRN for their dedication to duty and for daring to dream of a bright and wet future for Mars.

| CRISM ID | CRISM Mode | Earth DOY | Ls | % Coverage |
|---|---|---|---|---|
| Mars Year 28 | | | | |
| 2F70 | FRT | 2006 315 | 133 | ~50 |
| 3037 | MSP | 2006 320 | 136.4 | 0 |
| 3F8E | FRT | 2006 340 | 146.43 | 25 |
| 3B9C | MSP | 2007 003 | 160.38 | 90 |
| 7F89 | MSP | 2007 273 | 323 | 90 |
| Mars Year 29 | | | | |
| 8E90 | MSP | 2007 357 | 7 | 10 |
| 92AB | FRT | 2008 003 | 12 | 50 |
| 9474 | FRT | 2008 008 | 14.6 | 25 |
| 9654 | FRT | 2008 013 | 17 | 50 |
| 99FE | FRT | 2008 025 | 22.37 | 25 |
| 9C9C | FRT | 2008 036 | 27.61 | 95 |
| A266 | FRT | 2008 058 | 37.92 | 90 |
| A3BE | FRT | 2008 064 | 40.24 | 25 |
| A84E | FRT | 2008 109 | 60.25 | 25 |
| A928 | HRS | 2008 114 | 62.51 | 75 |
| AB68 | FRT | 2008 125 | 67.44 | 90 |
| B4B1 | FRT | 2008 182 | 92.20 | 95 |
| B965 | MSP | 2008 198 | 99.6 | 91 |
| D25A | MSP | 2008 303 | 149.46 | 99 |
| Mars Year 30 | | | | |
| 152F1 | MSP | 2010 002 | 32.46 | 80 |
| 166A4 | HRL | 2010 040 | 49.55 | 0 |
| 16FDB | FRT | 2010 063 | 59.54 | 95 |
| 173B2 | MSP | 2010 074 | 64.46 | 0 |
| 17994 | FRT | 2010 090 | 71.44 | 95 |
| 185A3 | HRL | 2010 118 | 83.34 | 90 |
| 18879 | FRT | 2010 124 | 86.05 | 75 |
| 18CCF | HRS | 2010 135 | 91.03 | 90 |
| 196E6 | HRS | 2010 173 | 108.05 | 75 |
| 1B7F0 | MSP | 2010 294 | 167.91 | 90 |
| 1C6FD | HRS | 2010 350 | 199.43 | 50 |
| Mars Year 31 | | | | |
| 232E0 | MSP | 2012 051 | 73.20 | 0 |
| 233FB | FRT | 2012 055 | 74.94 | 65 |
| 237CE | HRL | 2012 067 | 79.75 | 90 |
| 242CD | FRT | 2012 094 | 91.62 | 65 |
| 263D8 | MSP | 2012 193 | 137.94 | 3 |
| Mars Year 32 | | | | |
| 2ED28 | MSP | 2014 086 | 108.20 | 60 |
| 314FD | MSP | 2014 202 | 165.24 | 40 |

Table 1 – CRISM observations of Louth crater used in this study. The CRISM modes (which are discussed more fully in the text) are FRT= Full Resolution Targeted, HRL = Half Resolution Targeted, HRS = Half Resolution Short, MSP = Multispectral mapping. DOY = Day of Year.

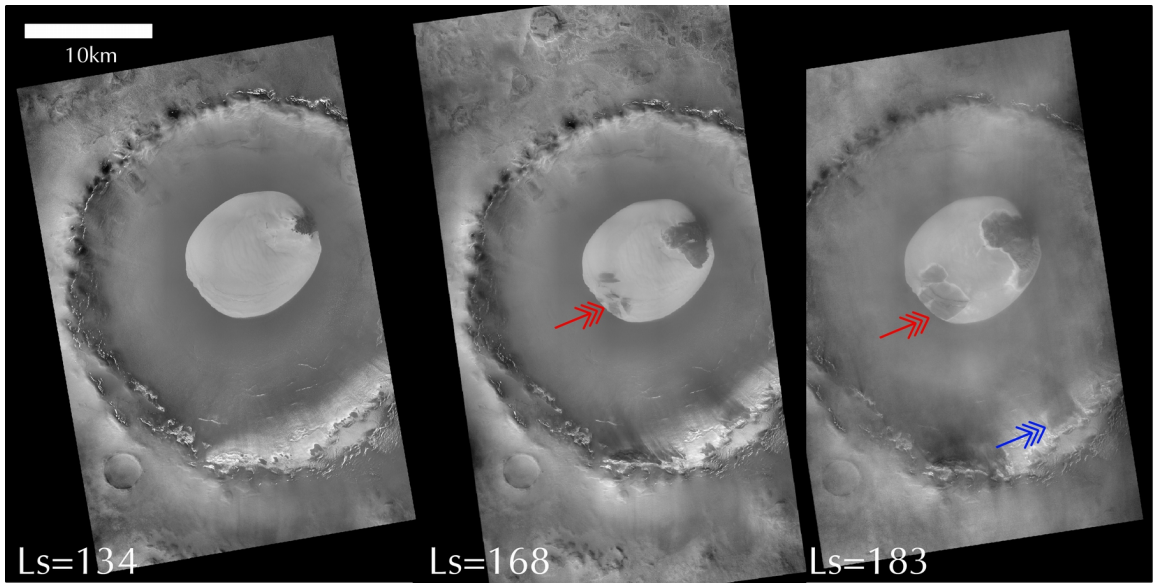

Figure 1 - Louth Crater as captured by MRO MSSS Context (CTX) camera in Mars Year 30 at $L_s$=134.1, 167.8 and 183.2, showing the development of an unexplained dark pattering (red arrows) in late summer to early-fall. North is up. The CTX instrument image identifiers are G02_019013_2503_XN_70N257W, G04_019857_2503_XN_70N257W, G05_020213_2503_XN_70N256W. The dark markings are largely limited to the ice mound, however some have developed on the ice covered pole-facing rim (blue arrows).

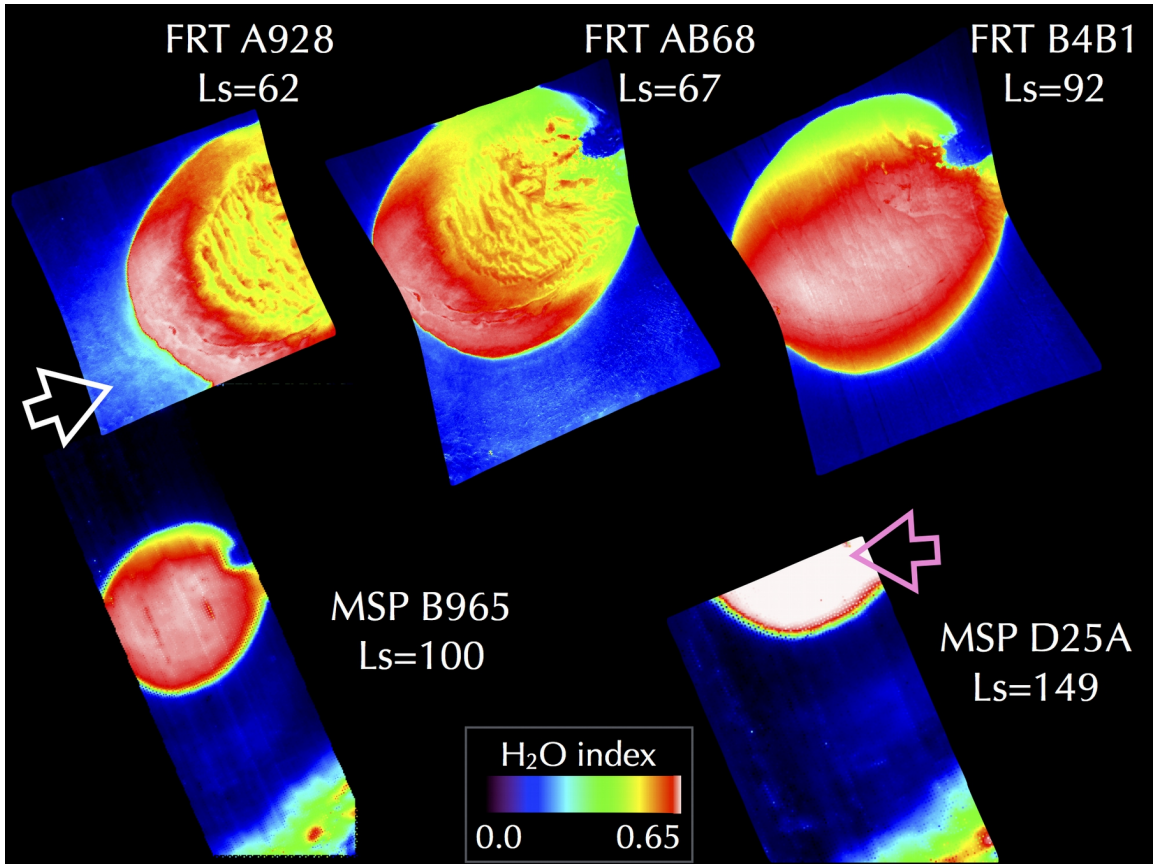

Figure 2a - Mars Year 29 CRISM observations of the evolution of the $H_2O$ index for late spring and early summer for Louth crater ice mound and the surrounding regolith. North is up. At $L_s$=62, water ice is seen on the regolith surrounding the water ice mound (white arrow). High $H_2O$ indexes correspond to larger grained water ice, primarily seen in the interior of the ice mound, although by $L_s$=149, the water ice index is uniformly high across the water ice mound (pink arrow), indicating the presence of relatively large grained water ice across the mound by that late summer period. The blue "bite" out of the north east corner of the ice mound corresponds to the ice-poor dark sand dunes at this location (see also Figure 1).

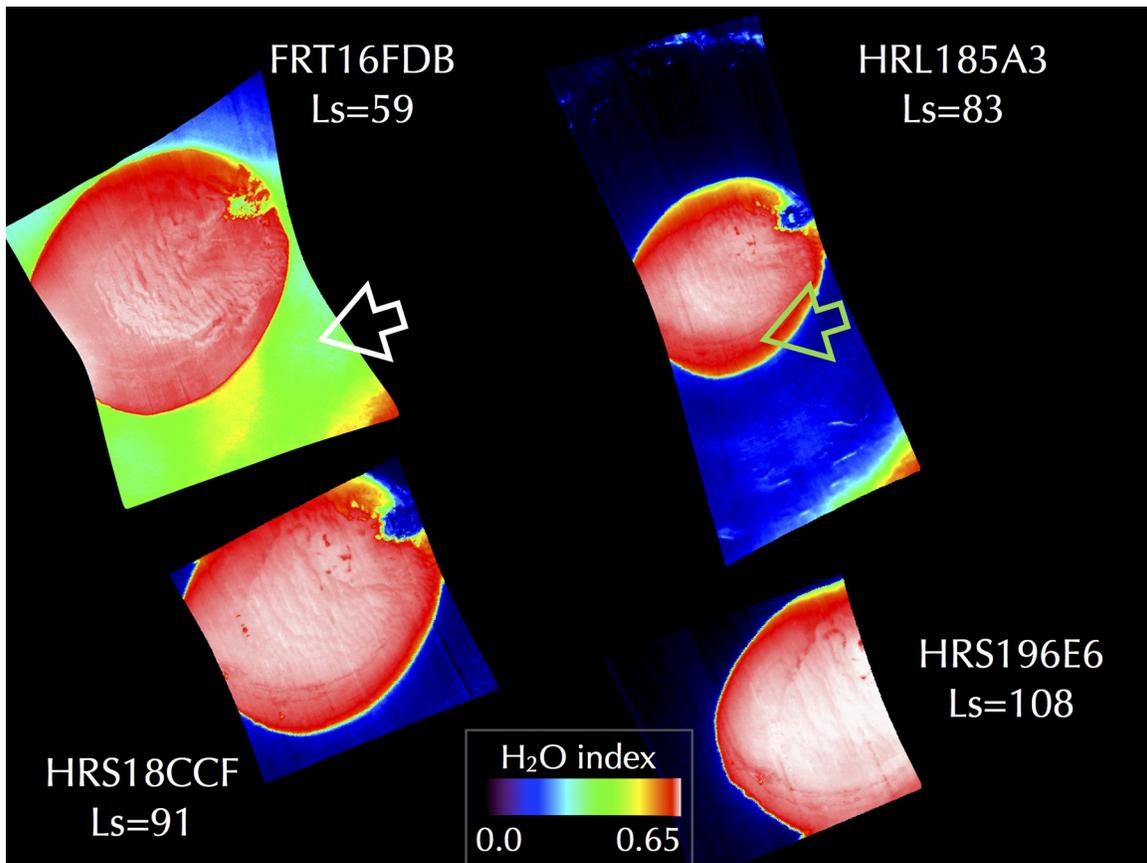

Figure 2b – Mars Year 30 $H_2O$ ice index maps of Louth Crater for northern spring and summer. North is up. At $L_s$=59, there is a large amount of water ice on the regolith around the water ice mound (white arrow). At $L_s$=83, the $H_2O$ ice index decreases at the edge of the water ice mound (this region is red and indicated by a green arrow). At $L_s$=91 and 108, there is a relatively uniform high water ice index across the ice mound.

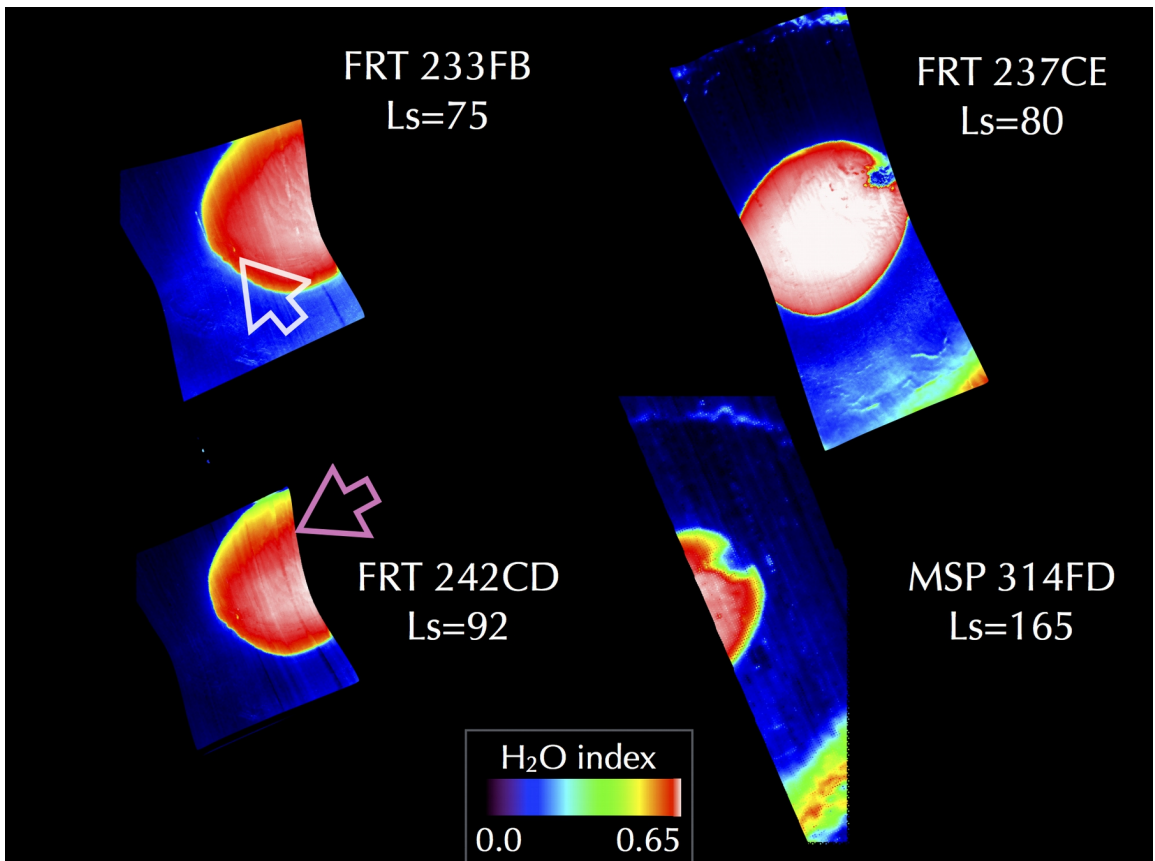

Figure 2c – Mars Year 31 $H_2O$ ice index maps of Louth Crater for northern spring and summer. North is up. At $L_s$=75, there is a relative low on the north edge of the water ice mound (this area is colored red and indicated by a white arrow). At $L_s$=80, the mound is relatively high and uniform across the mound (this is indicated in white). At $L_s$=92, the water ice index has decreased again on the north side of the ice mound (indicated by the pink arrow), for which we do not have a satisfactory explanation. At $L_s$=165 there is insufficient coverage of the mound to make firm conclusions regarding the water ice index spatial distribution across the mound. The water ice index of the crater rim (south east of the mound) is seen to be moderate in this image (it is colored red, yellow and green).

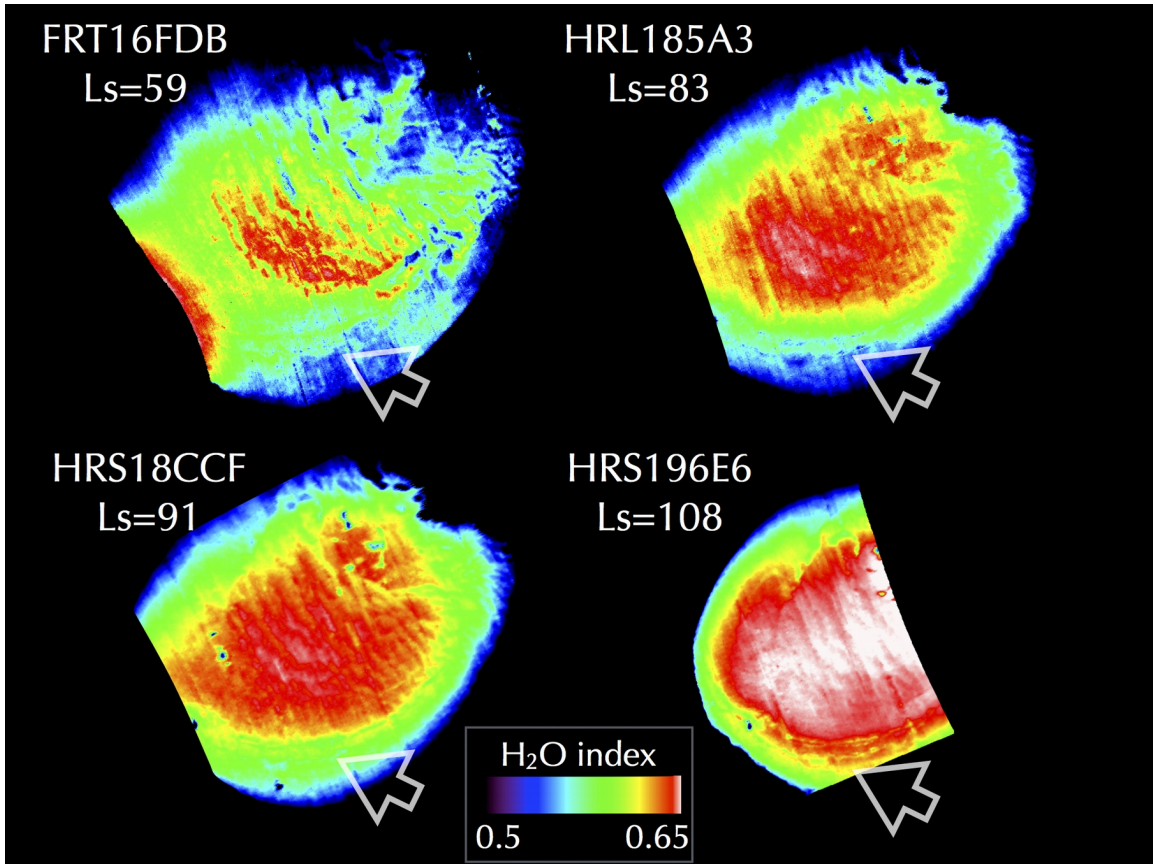

Figure 3 – $H_2O$ index image for Louth Crater for MY 30. This image is the same as Figure 2b, but stretch is altered to show the relative difference of the $H_2O$ ice index within the water ice mound. North is up. The $H_2O$ index of central part of the mound and the southern edge (arrows) are seen to brighten significantly from $L_s$=59-108.

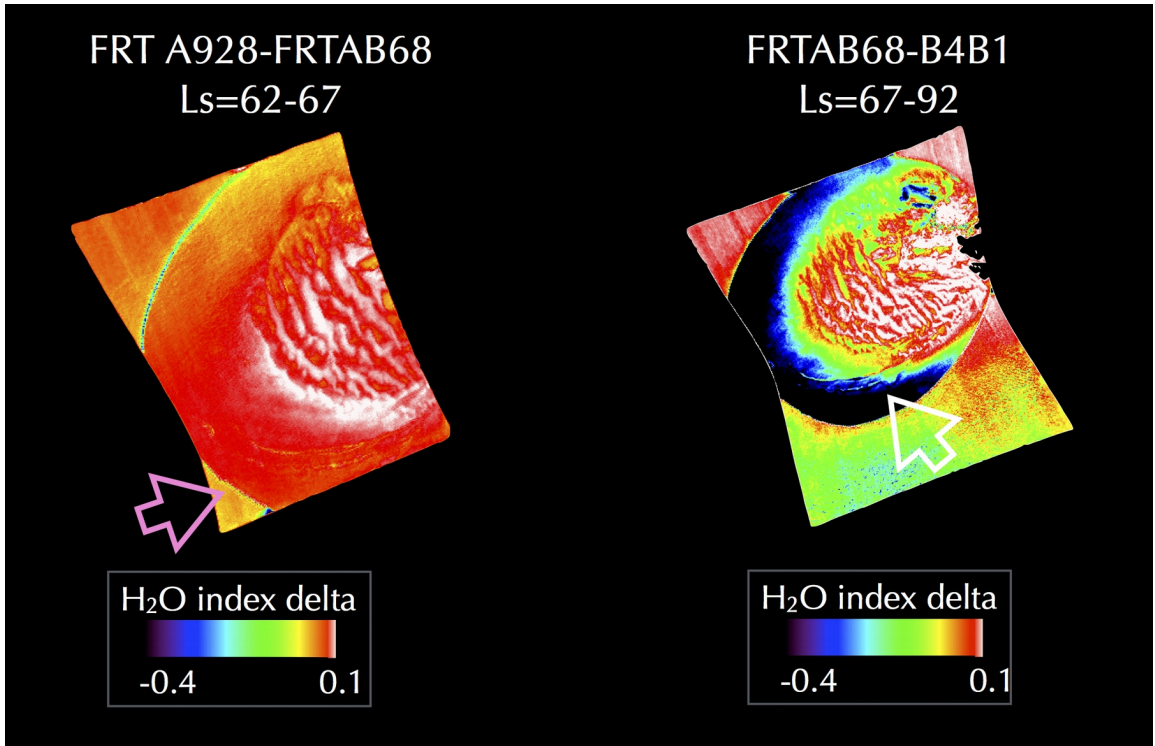

Figure 4a - Difference images for three images of Figure 3a, showing change in $H_2O$ index over summer for water ice mound and surrounds. North is up. Note the small decrease in water ice in the regolith from $L_s$=62-67 (this is colored orange and indicated by a pink arrow) and the large decrease in the $H_2O$ ice index on the edge of the mound from $L_s$=67-92 (this is colored black and indicated by a white arrow). The decrease in the $H_2O$ ice index is due to deposition of fine grained water ice, which has the effect of decreasing the depth of the $H_2O$ ice absorption band. See the text for further details.

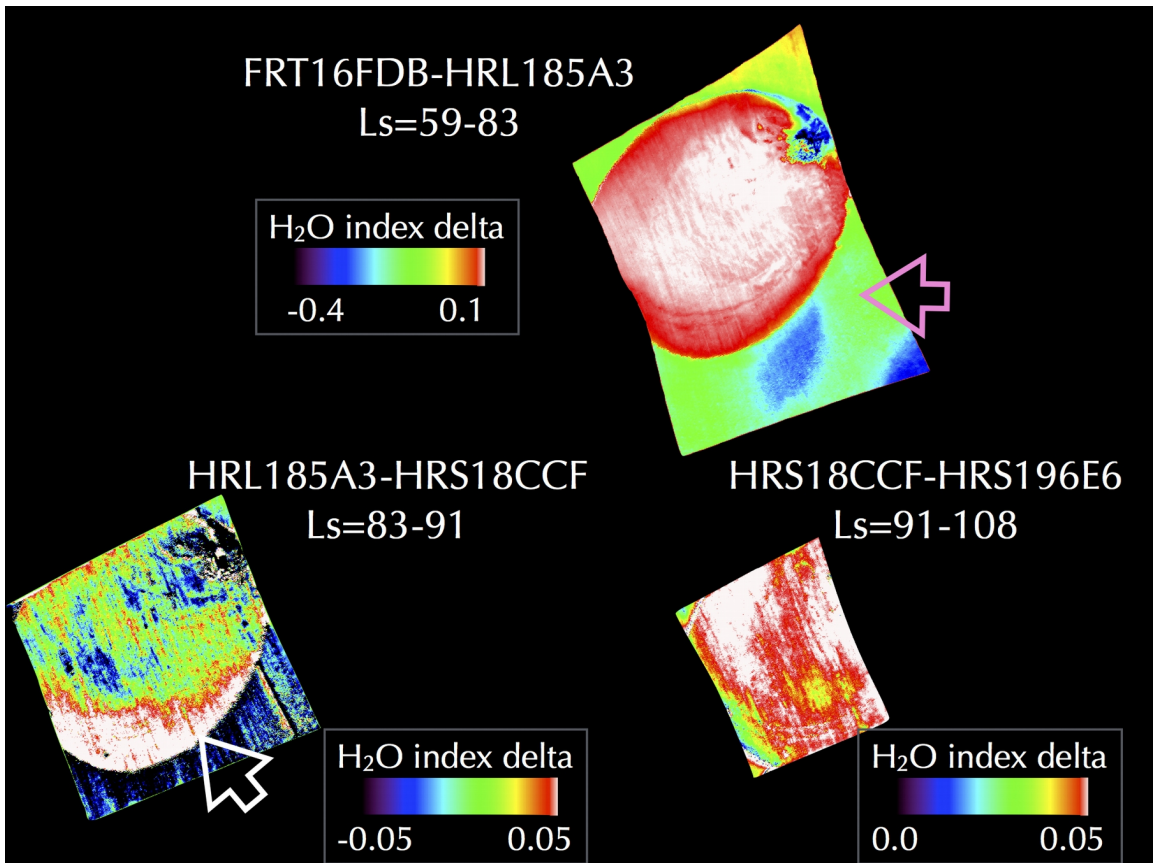

Figure 4b - Difference between four images in Figure 3b, showing change in $H_2O$ index over summer for water ice mound and surrounds. North is up. Note in $L_s$=59-83, there is a decrease in $H_2O$ ice index in the regolith surrounding the ice mound (pink arrow), particularly on the south side of the mound. Note also in $L_s$=83-91 there is a corresponding increase in the $H_2O$ index on the south side of the ice mound (grey arrows).

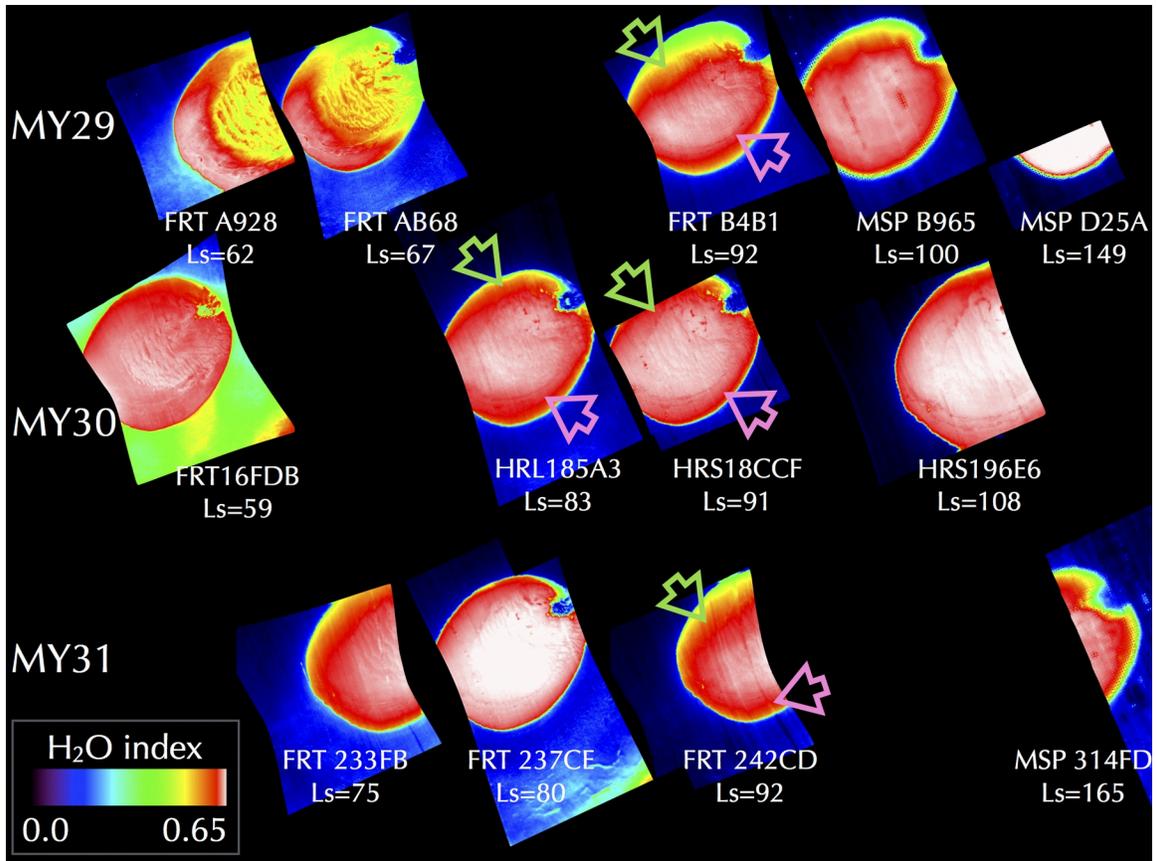

Figure 5 – $H_2O$ ice index images at Louth crater as a function of solar longitude (from Ls=59-149, left to right) for Mars Year 29, 30 and 31. Note that the temporal scale is not linear, the images are time ordered from left to right. Points to note: 1.) The amount of water ice on the surrounding regolith decreases from $L_s$=59 to $L_s$=165, and reaches a stable minimum amount by $L_s$=92. 2.) The water ice index on the water ice mound increases relatively steadily over the first two Mars Years (29, 30), however in one $L_s$=80 observation in Mars Year 31, the water ice index increases then decreases in $L_s$=92. This phenomenon is not as strongly apparent in the other Mars Years. 3.) There is a decease in the water ice index around the edge of the cap, particularly on the north side around $L_s$=92 in each Mars Year (green arrows) and also on the south side of the ice mound (pink arrows).

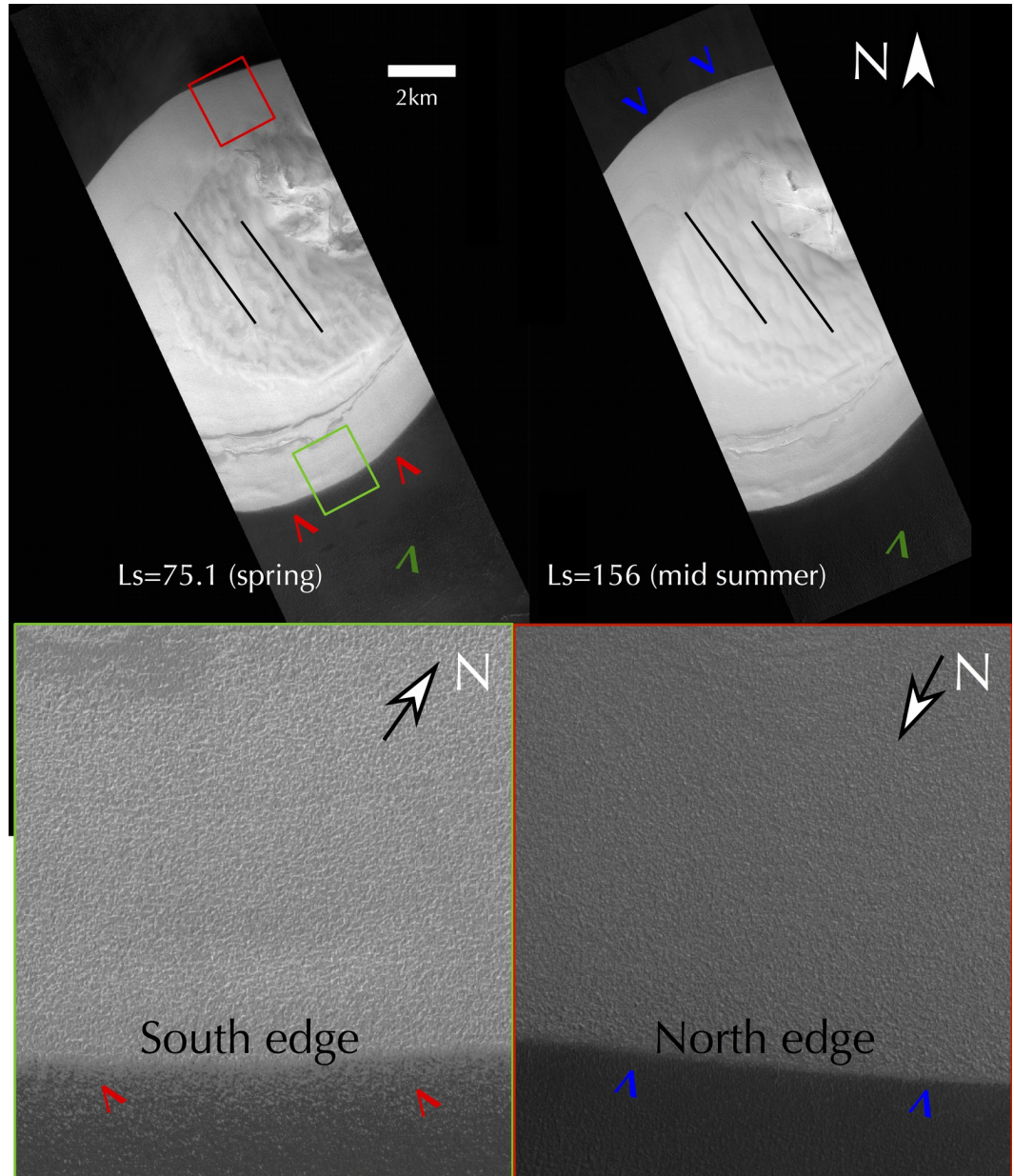

Figure 6 – HiRISE images of Louth crater during spring (PSP_008530, $L_s$=75.1, left) and mid-summer (ESP_037210, $L_s$=156, right). The close up images are from PSP_008530 alone to eliminate seasonal differences. The close up images are shown as boxes in the larger image, showing the rough stucco texture of the exterior rim of the water ice mound. The images show the northern rim of the ice mound contact with the regolith is clean and relatively distinct (blue arrows). The southern rim of the ice mound is ragged and less distinct (red arrows). This is a strong indication of southerly winds transporting water ice in the northerly direction, and is in accord with the asymmetric offset of the ice mound within Louth. The direction of this aeolian activity supports our hypothesis that winds are responsible for the asymmetry of the ice mound in the crater. The green arrows indicate a noticeable darkening of the regolith around the southern rim of the crater in mid summer (right), relative to the spring image (left). This supports

our hypothesis that fine grained (bright) water ice is lying on the regolith during spring, subliming during summer, and re-depositing onto the mound (see Figure 8). The linear features in the middle of the central ice mound (black lines) run almost north-south, which is in accord with their interpretation as wind blown sastrugi, which on Earth run parallel to the wind direction.

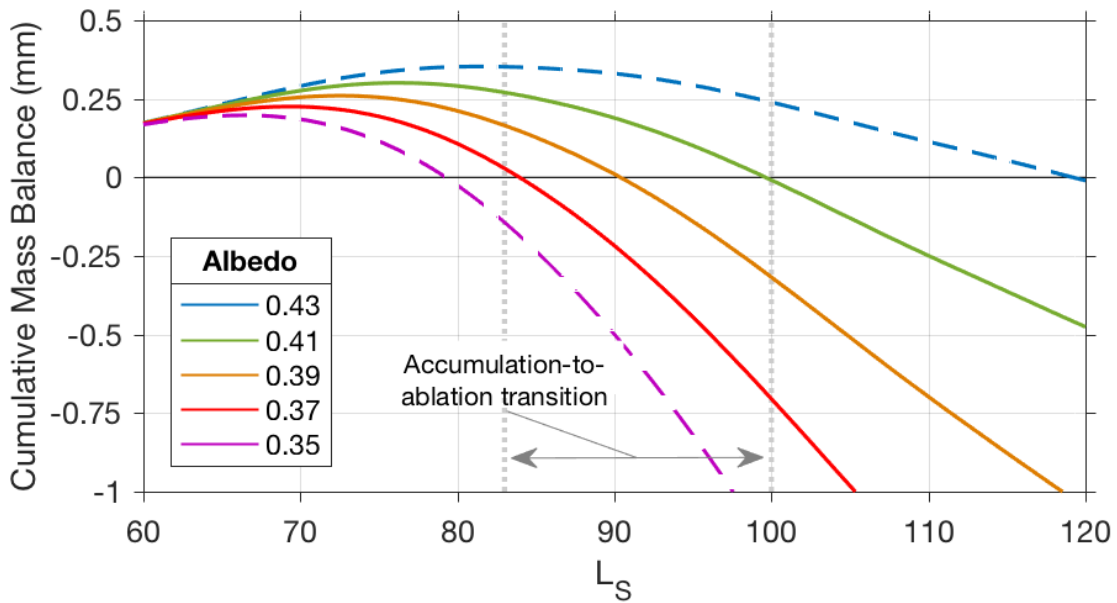

Figure 7 – Cumulative mass balance plot for a 1-D accumulation model for the water ice mound at Louth (see model description in Bapst et al. (2017)). The model assumes a constant windspeed of 3 m/s and is run for a range of albedo of 0.35-0.43. CRISM observations constrain the timing in the transition from accumulation to ablation (where zero is crossed in this plot) to $L_s$=83-100, which is consistent with modeled albedo of 0.37-0.41 (solid lines).

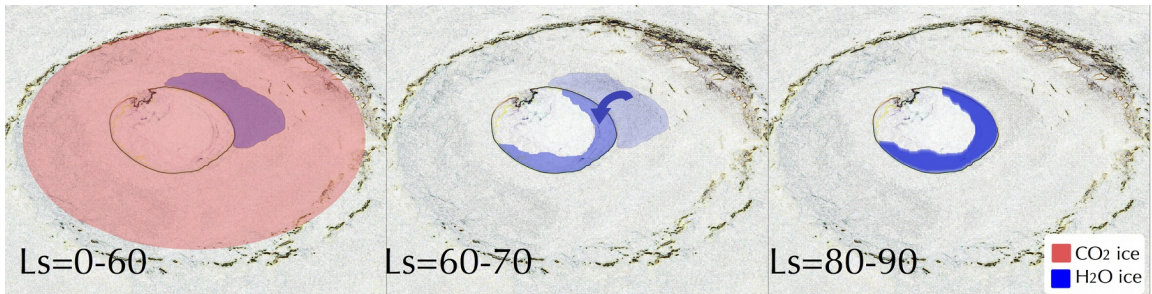

Figure 8 – Diagram of $H_2O$ transfer process at Louth Crater during northern spring and summer. (left) From $L_s$=0-60, Louth Crater is covered in $CO_2$ ice, shown in red. Water ice (blue) in the regolith is trapped. (center) At $L_s$=60, the $CO_2$ ice has sublimed and water ice in the regolith begins to sublime, and is trapped on the nearby cold surface of the water ice mound, depositing a find grained layer around the smooth edge of the ice mound. (right) From $L_s$=80-90, no more water ice is available on the regolith around the ice mound, and the ice around the boundary of the mound is fine grained and relatively smooth.